\documentclass[11pt]{article}

\usepackage{amsmath}
\usepackage{graphicx}
\usepackage{amsfonts}
\usepackage{amssymb}
\usepackage{epsfig}
\usepackage{color}
\usepackage{psfrag}

\newcommand{\EQ}{\begin{equation}}
\newcommand{\EN}{\end{equation}}

\begin{document}
\setcounter{page}{0} \topmargin 0pt
\renewcommand{\thefootnote}{\arabic{footnote}}
\newpage
\setcounter{page}{0}

\begin{titlepage}

\begin{flushright}
ISAS/44/2004/FM
\end{flushright}

\vspace{0.5cm}
\begin{center}
{\Large {\bf Semiclassical Energy Levels of Sine--Gordon Model}}\\
{\Large {\bf  on a Strip with Dirichlet Boundary Conditions}}

\vspace{2cm}
{\large G. Mussardo$^{1,2}$, V. Riva$^{1,2}$ and G. Sotkov$^{3}$} \\
\vspace{0.5cm} {\em $^{1}$International School for Advanced Studies}\\
{\em Via Beirut 1, 34100 Trieste, Italy} \\
\vspace{0.3cm} {\em $^{2}$Istituto Nazionale di Fisica Nucleare}\\
{\em Sezione di Trieste}\\\vspace{0.3cm} {\em $^3$ Institute for Nuclear Research and
Nuclear Energy} \\
{\em Bulgarian Academy of Sciences} \\
{\em Tsarigradsko Chaussee 72, BG-1784, Sofia, Bulgaria}

\end{center}
\vspace{1cm}

\begin{abstract}
\noindent We derive analytic expressions of the semiclassical energy levels
of Sine--Gordon model in a strip geometry with Dirichlet boundary
condition at both edges. They are obtained by initially selecting
the classical backgrounds relative to the vacuum or to the kink
sectors, and then solving the Schr\"{o}dinger equations (of
Lam\`{e} type) associated to the stability condition. Explicit
formulas are presented for the classical solutions of both the
vacuum and kink states and for the energy levels at arbitrary
values of the size of the system. Their
ultraviolet and infrared limits are also discussed. \vspace{2cm}

\hrulefill

E-mail addresses: mussardo@sissa.it,\,\, riva@sissa.it,\,\,
sotkov@inrne.bas.bg

\end{abstract}

\end{titlepage}

\newpage

\section{Introduction}\label{intro}
\setcounter{equation}{0}
Since their introduction in the seminal works \cite{DHN,goldstone},
semiclassical methods have proved to be efficient tools for analysing
non-perturbative effects in a large class of quantum field theories. Based on this approach,
there have been recently new developments concerning form factors at
a finite volume \cite{finvolff}, non--integrable models \cite{dsgmrs}
and energy levels of a quantum field theory on a cylinder geometry
\cite{SGscaling}. As we show in the following, the analysis done in
\cite{SGscaling} also admits an interesting generalization to a
quantum field theory defined on a strip of width $R$, with certain
boundary conditions at its edges. The example discussed here is the
Sine--Gordon model, defined by the Lagrangian
\begin{equation}\label{lagr}
{\cal L}\,=\,\frac{1}{2}\partial_\mu \phi\,\partial^\mu \phi
- \frac{m^{2}}{\beta^{2}} \left(1-\cos\beta\phi
\right)\,\,\,,
\end{equation}
subjected to the Dirichlet boundary conditions (D.b.c.)
\begin{equation}\label{Dbc}
\phi(0,t)\,=\,\phi_{0}+\frac{2\pi}{\beta}\,n_{0}\;,\qquad
\phi(R,t)\,=\,\phi_{R}+\frac{2\pi}{\beta}\,n_{R}\;,\qquad \forall
t\;
\end{equation}
with $0\leq \phi_{0,R}< \frac{2\pi}{\beta}$ and $n_{0,R}\in \mathbb{Z}$. The
topological charge of this model is conserved also in the presence of boundaries
and it can be conveniently defined as
\begin{equation}
Q\,\equiv\,\frac{\beta}{2\pi}
\left\{\int\limits_{0}^{R}\partial_{x}\phi\,dx\,-\,(\phi_{R}-\phi_{0})\right\}\,=\,
n_{R}-n_{0}\;.
\end{equation}
Hence the space of states is split in topological
sectors with $Q=0,\pm 1,\pm 2...$, and within a given $Q$-sector the states are
characterized by their energies only.

It is worth mentioning that, in recent years, this problem (and variations thereof)
has attracted the attention of several groups: the case
of half--plane geometry, for instance, has been discussed by bootstrap
methods in \cite{ghoshzam,mattdorey,BPTT} and by semiclassical ones
in \cite{SSW,corr,kormospalla} whereas the thermodynamics of different cases
in a strip geometry has been studied in a series of publications (see [12--18]).
%\cite{SS,LMSS,cauxsal,leerim,rim,ahnnep,rav}.
%In completing this work, a paper on a (semi)classical analysis of
%Sine--Gordon model on a strip \cite{hungstrip} also appeared,
%which partially overlaps with ours.

The semiclassical quantization presented here adds new pieces of
information on this subject and it may be seen as complementary to
the aforementioned studies: for the static solutions, it basically
consists of identifying, in the limit $\beta \rightarrow 0$, a
proper classical background $\phi_{cl}(x)$ for the given sector of
the theory in exam, and then expressing the semiclassical energy
levels as
\begin{equation}\label{DHNlevels}
E_{\{k_n\}}\,=\,{\cal E}_{cl}\,+\,\sum\limits_{n}
\left(k_n+\frac{1}{2}\right)\omega_n\;,\qquad k_n\in
\mathbb{N}\;,
\end{equation}
where ${\cal E}_{cl}$ is the classical energy of the solution whereas
the frequencies $\omega_n$ are the eigenvalues of the so-called
\lq\lq stability equation'' \cite{DHN}
\begin{equation}\label{DHNstability}
\left[-\frac{d^2}{dx^2}+V''(\phi_{cl})\right]\,\eta_n(x)\,=\,
\omega_n^2\,\eta_n(x)\;.
\end{equation}
For the Sine--Gordon model with periodic boundary conditions,
alias in a cylinder geometry, this program has been completed in
\cite{SGscaling}. Given the similarity of the outcoming formulas
with the ones appearing in \cite{SGscaling}, in the sequel we will
often refer to that paper for the main mathematical definitions as
well as for the discussion of some technical details. There is
though a conceptual difference between the periodic example
and the one studied here: in the
periodic case, in fact, the vacuum sector is trivial at the
semiclassical level (it simply corresponds to the constant
classical solution) and therefore the  semiclassical quantization
provides non-perturbative results just starting from the one-kink
sector. Contrarily, on the strip with Dirichlet b.c., the vacuum
sector itself is represented by a non-trivial classical solution
and its quantization is even slightly more elaborated than the one 
of the kink sectors.

\section{Static classical solutions}
\setcounter{equation}{0}

\begin{figure}[h]
\begin{tabular}{p{8cm}p{8cm}}

\footnotesize

\psfrag{phicl(x)}{$\beta\phi^{+}_{cl}$} \psfrag{2 pi}{$2\pi$}
\psfrag{K(k^2)}{$\textbf{K}(k^{2})$}
\psfrag{-K(k^2)}{$-\textbf{K}(k^{2})$}
\psfrag{x}{$\hspace{0.2cm}u^{+}$}

\psfig{figure=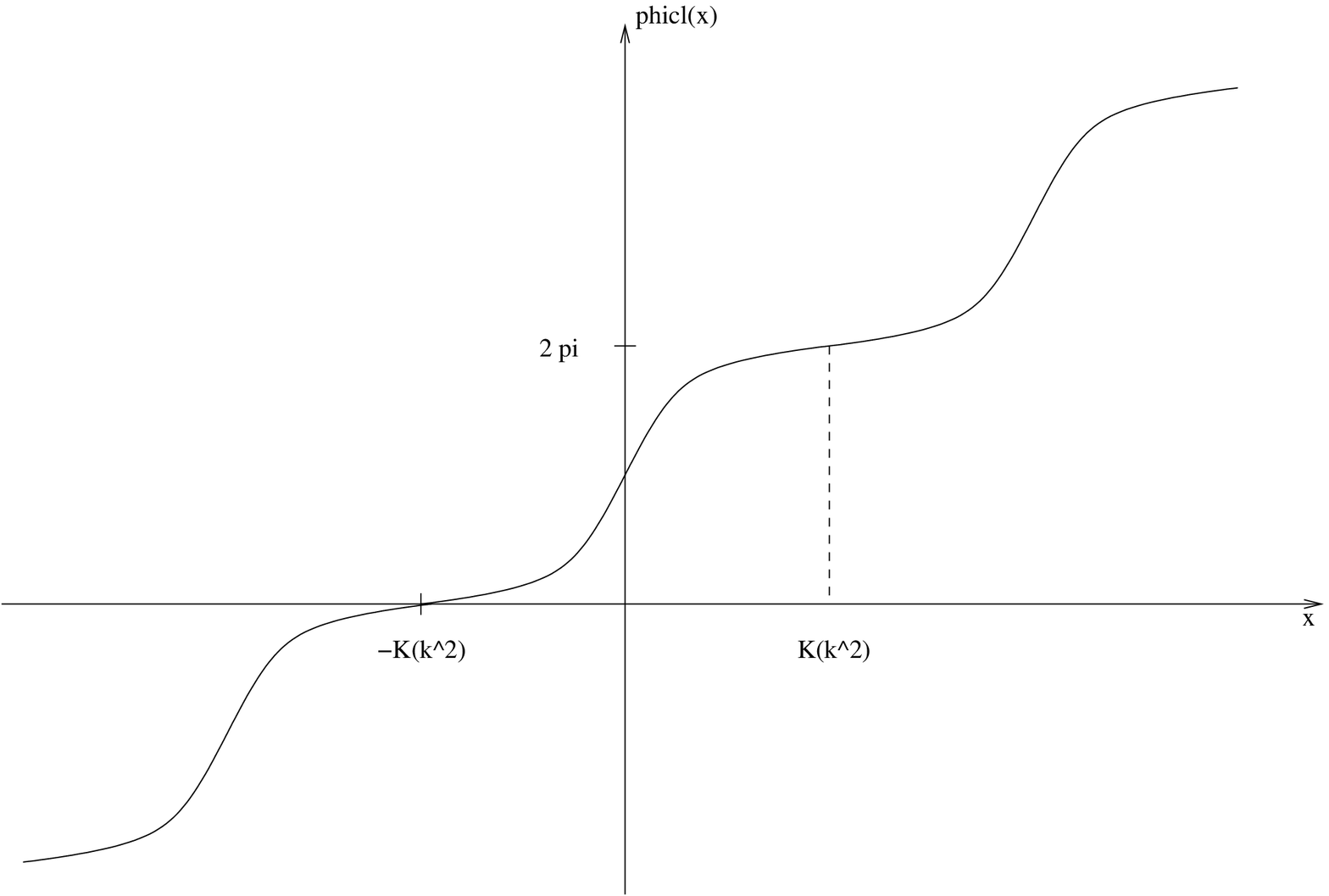,height=5cm,width=7cm}&

\footnotesize

\psfrag{phicl(x)}{$\beta\phi^{-}_{cl}$}
\psfrag{phi0}{$\beta\tilde{\phi}$} \psfrag{2
pi-phi0}{$2\pi-\beta\tilde{\phi}$}
\psfrag{K(k^2)}{$\textbf{K}(k^{2})$}
\psfrag{-K(k^2)}{$-\textbf{K}(k^{2})$}
\psfrag{x}{$\hspace{0.2cm}u^{-}$}

\psfig{figure=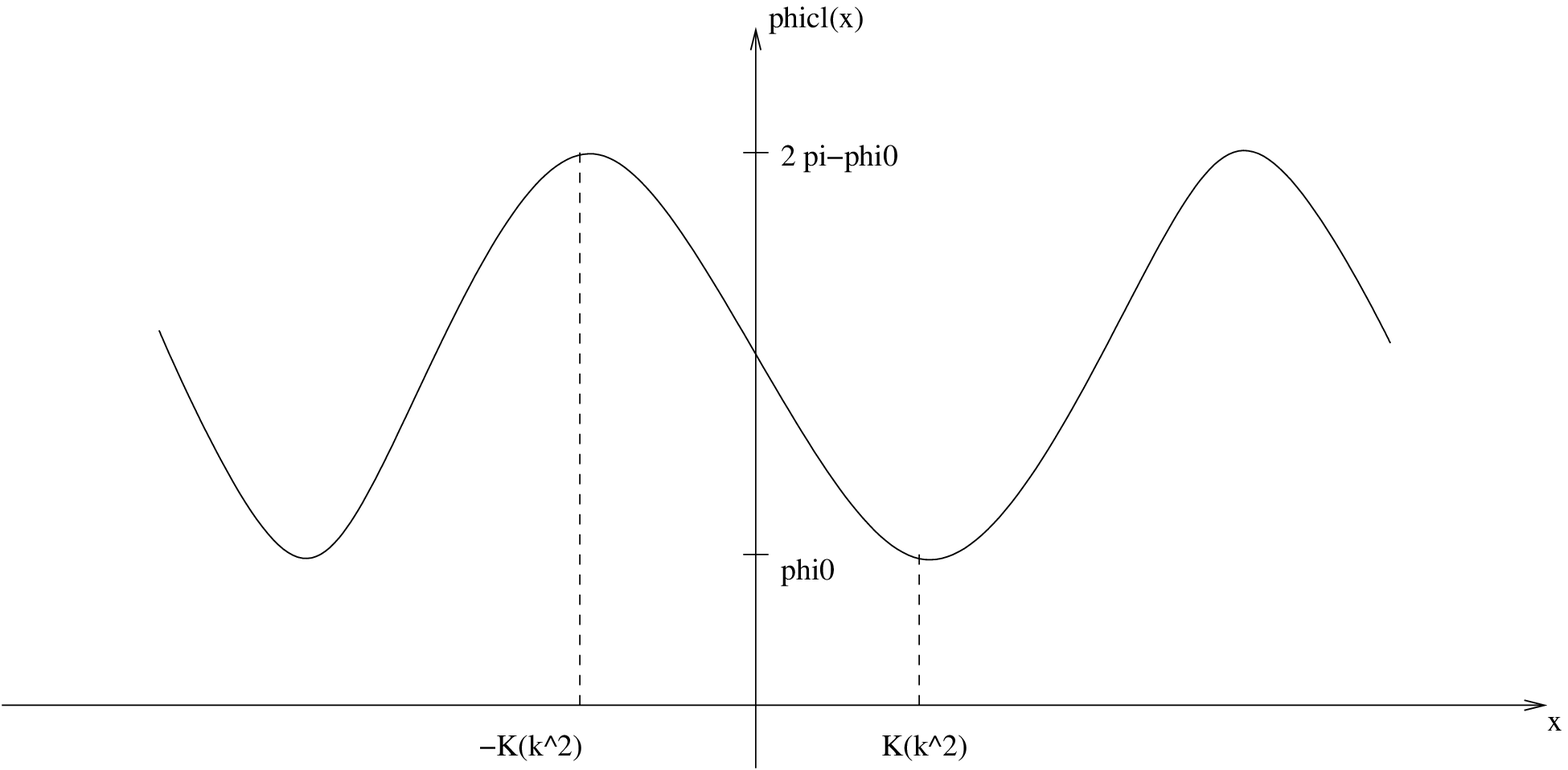,height=5cm,width=7cm}
\end{tabular}
\caption{Solutions of eq.\,(\ref{firstSG}), $A > 0$ (left hand
side), $-2 < A < 0$ (right hand side).}
 \label{figSGsol}
\end{figure}

In the static case, the Euler--Lagrange equation of motion associated to
(\ref{lagr}) is equivalent to the first order differential equation
\begin{equation}
\frac{1}{2}\left(\frac{\partial \phi_{cl}} {\partial x}\right)^{2}
\,=\, \frac{m^{2}}{\beta^{2}} \left(1-\cos\beta\phi_{cl} + A
\right) \;, \label{firstSG}
\end{equation}
which admits three kinds of solution, depending on the sign of
the constant $A$. The simplest corresponds to $A=0$ and it
describes the standard kink in infinite volume:
$$
\phi^{0}_{cl}(x)\,=\,\frac{4}{\beta}\,\arctan\,e^{ m(x-x_{0})}\;.
$$
In this paper, we will be concerned with the solutions relative to
the case $A\neq 0$, which can be expressed in terms of Jacobi
elliptic functions\footnote{See \cite{SGscaling} and references
therein for the definitions and the basic properties of complete
elliptic integrals $\textbf{K}(k)$ and $\textbf{E}(k)$, 
Jacobi and Weierstrass functions. In the following
we will also use the incomplete elliptic integrals
\begin{equation*}\label{ellint}
F(\varphi,k)\,=\,\int\limits_{0}^{\varphi}\frac{d\alpha}
{\sqrt{1-k^{2}\sin^{2}\alpha}}\;,\qquad
E(\varphi,k)\,=\,\int\limits_{0}^{\varphi}d\alpha
\sqrt{1-k^{2}\sin^{2}\alpha}\;,
\end{equation*}
which reduce to the complete ones at $\varphi=\pi/2$.}
\cite{takoka}. In particular, for $ A > 0$ we have
\begin{equation}
\phi^{+}_{cl}(x)\,=\,\frac{\pi}{\beta} + \frac{2}{\beta}\,
\textrm{am}\left(\frac{ m (x - x_0)}{k},k\right)\;, \qquad
k^{2} \,=\,\frac{2}{2+A}\,\,\,, \label{SGam}
\end{equation}
which has the monotonic and unbounded behaviour in terms
of the real variable $u^{+}=\frac{ m (x - x_0)}{k}$ shown in
Fig.\,\ref{figSGsol}. For $ -2<A < 0$, the solution is given
instead by
\begin{equation}
\phi^{-}_{cl}(x)\,=\, \frac{2}{\beta}\,
\arccos\left[k\;\textrm{sn}\left( m (x -
x_0),k\right)\right]\;, \qquad k^{2} \,=\,1+\frac{A}{2}\,\,\,,
\label{SGsn}
\end{equation}
and it oscillates in the real variable $u^{-}= m (x - x_0)$
between the $k$-dependent values $\tilde{\phi}$ and
$\frac{2\pi}{\beta}-\tilde{\phi}$ (see Fig.\,\ref{figSGsol}).

The SG model with the Dirichlet b.c. (\ref{Dbc}) can be classically described
by using the two building functions $\phi^{+}_{cl}(x)$ and $\phi^{-}_{cl}(x)$,
thanks to their free parameters $x_0$ and $k$, which can be fixed in terms of 
$\phi_0$, $\phi_R$ and $R$. However, in order to simplify the notation, in writing
down our solutions we will rather use $R$ and $x_0$, both considered 
as functions of $\phi_0$, $\phi_R$ and $k$ 
(as a matter of fact, $k$ can be recovered by inverting the elliptic integrals 
which enter the corresponding expression of $R$).  

As shown below, both types of solutions $\phi^{+}_{cl}(x)$ and $\phi^{-}_{cl}(x)$ 
are needed, in general, to define the classical background in the vacuum sector whereas 
only one of them, $\phi^{+}_{cl}(x)$, is employed for implementing the Dirichlet b.c. 
in the kink sector.

\section{The vacuum sector: $Q=0$}
\setcounter{equation}{0}

To discuss the vacuum sector, it is sufficient to restrict the
attention to the case\footnote{All other cases can be described in
a similar way, defining properly $x_{0}$ and $R$, and by using
antikinks when necessary.} $n_0=n_R=0$, $\phi_{0}<\phi_{R}$ and
$\left|\cos\frac{\beta}{2}\phi_{0}\right|>\left|\cos\frac{\beta}{2}\phi_{R}\right|$.
It is also convenient to introduce the compact notation
\begin{equation}
c\,_{0,R} \,\equiv\, \cos\frac{\beta}{2}\,\phi_{0,R}\;\;.
\end{equation}

\begin{figure}[h]
\begin{tabular}{p{4.5cm}p{5.5cm}p{6.5cm}}

\footnotesize

\psfrag{phi}{$\beta\phi^{(1)}_{cl}$}
\psfrag{ell}{$x$}

\psfig{figure=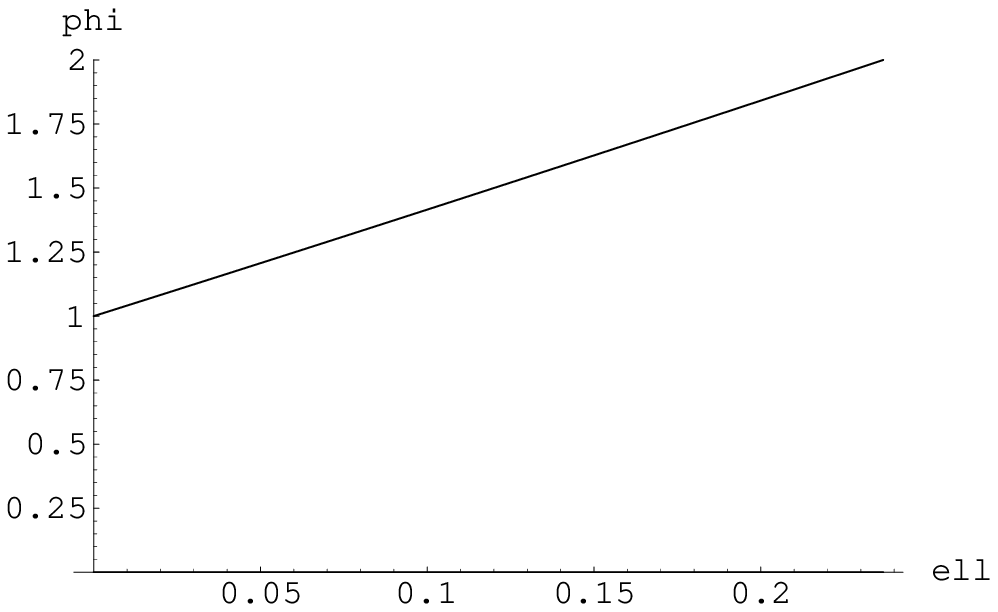,height=4cm,width=4cm}&

\footnotesize

\psfrag{phi}{$\beta\phi^{(2)}_{cl}$}
\psfrag{ell}{$x$}

\psfig{figure=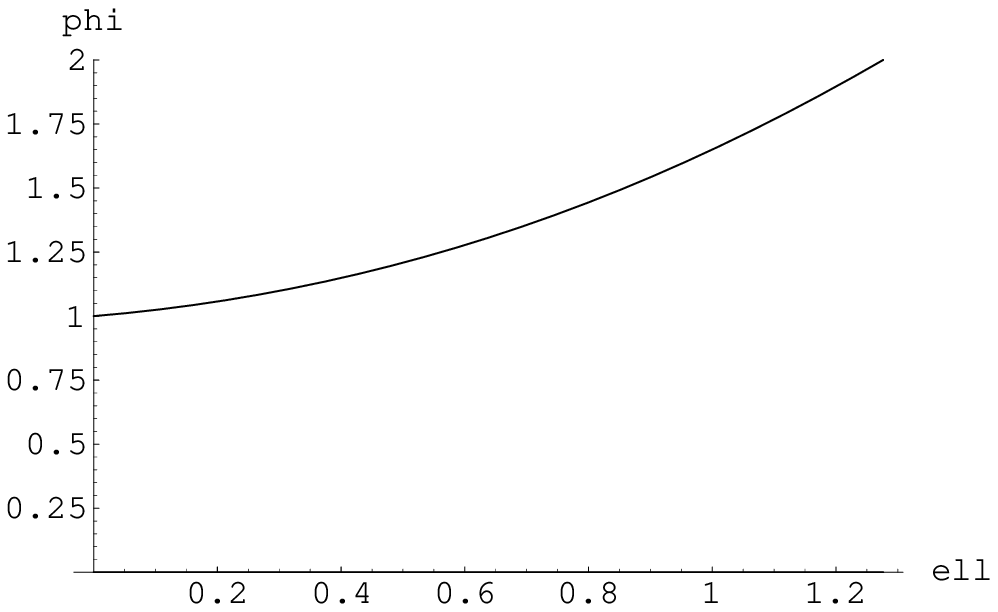,height=4cm,width=5cm}&

\footnotesize

\psfrag{phi}{$\beta\phi^{(3)}_{cl}$}
\psfrag{ell}{$x$}

\psfig{figure=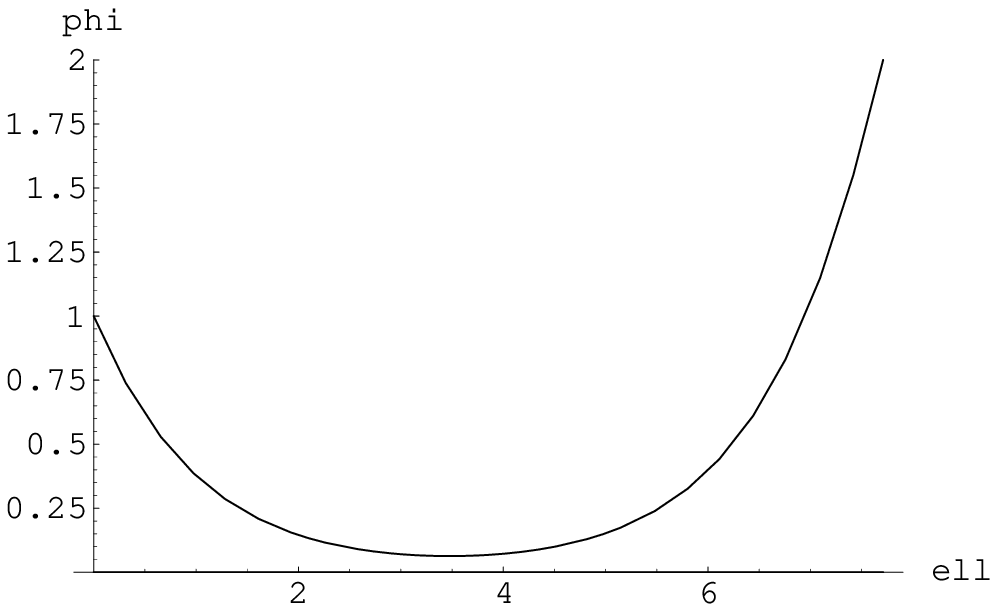,height=4cm,width=6cm}
\end{tabular}
\caption{Classical solution (\ref{vacD}) at some value of $R$, in the case $\beta\phi_{0}=1$
and $\beta\phi_{R}=2$.} \label{figvac}
\end{figure}

In order to write down explicitly the classical background
corresponding to the vacuum state with Dirichlet b.c., it is
necessary to introduce preliminarily two particular values $R_1$
and $R_2$ of the width $R$ of the strip, which mark a change in
the nature of the solution. They are given by
$$
\begin{cases}
mR_{1}\,=\,\text{arctanh}\left(c\,_{0}\right)-
\text{arctanh}\left(c_{R}\right)\;,\\
mR_{2}\,=\,\textbf{K}(\tilde{k})
-F\left(\arcsin\frac{c_R}{\tilde{k}^{}}
\,,\;\tilde{k}\right)\;,\hspace{1.5cm}
\tilde{k}=\left|\,c\,_{0}\right|\;.
\end{cases}
$$
With these definitions, the classical vacuum solution, as a function
of $x\in[0,R]$, has the following behaviour in the three regimes of $R$:
\begin{equation}\label{vacD}
\phi^{\text{vac}}_{cl}(x)\,=\,
\begin{cases}
\phi^{(1)}_{cl}(x)\qquad \text{for}\qquad 0<R<R_{1}\\
\phi^{(2)}_{cl}(x)\qquad \text{for}\qquad R_{1}<R<R_{2}\\
\phi^{(3)}_{cl}(x)\qquad \text{for}\qquad R_{2}<R<\infty
\end{cases}
\end{equation}
where
\begin{equation*}
\phi^{(1)}_{cl}(x)\,=\,\phi^{(+)}_{cl}(x)\qquad \text{with}\qquad
\begin{cases}
m x_{0} = -k\,F\left(\frac{\beta}{2}\,\phi_{0}-\frac{\pi}{2}\,,\;k\right)\\
m R =
k\left[F\left(\frac{\beta}{2}\,\phi_{R}-\frac{\pi}{2}\,,\;k\right)-
F\left(\frac{\beta}{2}\,\phi_{0}-\frac{\pi}{2}\,,\;k\right)\right]\\
 0 < k < 1 \end{cases}
\end{equation*}
\begin{equation*}\hspace{-1cm}
\phi^{(2)}_{cl}(x)\,=\,\phi^{(-)}_{cl}(x)\qquad \text{with}\qquad
\begin{cases}
mx_{0} = -2\textbf{K}(k)+F\left(\arcsin\frac{c\,_0}{k}\,,\;k\right)\\
m R = F\left(\arcsin\frac{c\,_0}{k}\,,\;k\right)-
F\left(\arcsin\frac{c_R}{k}\,,\;k\right)\\
\tilde{k} < k < 1\end{cases}
\end{equation*}
\begin{equation*}\hspace{0.5cm}
\phi^{(3)}_{cl}(x)\,=\,\phi^{(-)}_{cl}(x)\qquad \text{with}\qquad
\begin{cases}
mx_{0} = -F\left(\arcsin\frac{c\,_0}{k}\,,\;k\right)\\
m R =
2\textbf{K}(k)-F\left(\arcsin\frac{c\,_0}{k}\,,\;k\right)-
F\left(\arcsin\frac{c_R}{k}\,,\;k\right)\\
\tilde{k} < k < 1\end{cases}
\end{equation*}
It is easy to check that at the particular values $R_1$ and $R_2$, the different
definitions of the background nicely coincide. Fig.\,\ref{figvac} shows the
classical solution at some values of $R$, one for each of the three
regimes\footnote{We have chosen for the plot the specific values
$\beta\phi_0 = 1$ and $\beta\phi_R = 2$, for which $mR_1 = 0.76$ and
$mR_2 = 1.49$. The same values will be considered in all other pictures 
since their qualitative features 
do not sensibly depend on these parameters, except for few particular values of 
$\phi_{0,R}$ discussed separately}.

\vspace{5mm}

\begin{figure}[h]
\footnotesize \psfrag{ec}{$\beta^{2}{\cal E}_{cl}^{\text{vac}}/m$}
\psfrag{ell}{$mR$}
\begin{center}
\psfig{figure=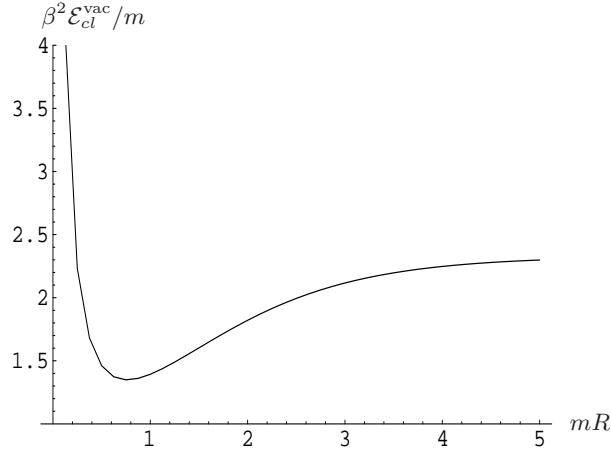,height=6cm,width=8cm}
\end{center}
\caption{Classical energy (\ref{eclD}) for $\beta\phi_{0}=1$
and $\beta\phi_{R}=2$.} \label{figecl}
\end{figure}

The classical energy of the background (\ref{vacD}) is expressed
as
\begin{equation}\label{eclD}
{\cal E}^{\text{vac}}_{cl}(R)\,=\,
\begin{cases}
{\cal E}^{(1)}_{cl}(R)\qquad \text{for}\qquad 0<R<R_{1}\\
{\cal E}^{(2)}_{cl}(R)\qquad \text{for}\qquad R_{1}<R<R_{2}\\
{\cal E}^{(3)}_{cl}(R)\qquad \text{for}\qquad R_{2}<R<\infty
\end{cases}
\end{equation}
where
\begin{eqnarray}
{\cal
E}^{(1)}_{cl}(R)&=&\frac{2m}{\beta^{2}}\left\{\left(1-\frac{1}{k^{2}}\right)m
R+\frac{2}{k}\left[E\left(\frac{\beta}{2}\,\phi_{R}-\frac{\pi}{2},k\right)-
E\left(\frac{\beta}{2}\,\phi_{0}-\frac{\pi}{2},k\right)\right]\right\}\;,\nonumber\\
{\cal E}^{(2)}_{cl}(R)&=&\frac{2m}{\beta^{2}}\left\{(k^{2}-1)m
R+2\left[E\left(\arcsin\frac{c\,_0}{k}\,,\;k\right)-
E\left(\arcsin\frac{c_R}{k}\,,\;k\right)\right]\right\}\;,\label{eclDexplicit}\nonumber\\
{\cal E}^{(3)}_{cl}(R)&=&\frac{2m}{\beta^{2}}\left\{(k^{2}-1)m
R+2\left[2\textbf{E}(k)-E\left(\arcsin\frac{c\,_0}{k}\,,\;k\right)-
E\left(\arcsin\frac{c_R}{k}\,,\;k\right)\right]\right\}\;,\nonumber
\end{eqnarray}
and it is plotted in Fig.\,\ref{figecl}. As expected, the quantity (\ref{eclD})
has a smooth behaviour at $R_{1}$ and $R_{2}$, which correspond to the minimum
and the point of zero curvature of this function, respectively. The non monotonic
behaviour of the classical energy gives an intuitive motivation for the classical
background being differently defined in the three regimes of $R$.

Furthermore, the classical energy can be easily expanded in the
ultraviolet (UV) or infrared (IR) limit, i.e. for small or large
values of $mR$, which correspond to $k\to 0$ in the regime
$0<R<R_1$ or to $k\to 1$ in the regime $R_2<R<\infty$,
respectively.

In fact, expanding the elliptic integrals in (\ref{eclD}) (see
\cite{GRA} for the relative formulas), and comparing the result
order by order with the small-$k$ expansion of $mR$ defined
in the first regime of (\ref{vacD})
\begin{equation}\label{mRUV}
mR\,=\,k\,\frac{\beta}{2}(\phi_R-\phi_0)\left[1+\frac{k^2}{4}\left(1+
\frac{\sin\beta\phi_R-\sin\beta\phi_0}{\beta(\phi_R-\phi_0)}\right)
+ \cdots \right]\;,
\end{equation}
one obtains the small-$mR$ behaviour
\begin{equation}\label{eclUV}
{\cal E}^{(1)}_{cl}(R)\,=\,\frac{1}{2 R}(\phi_R-\phi_0)^2+R\,\frac{m^2}{\beta^2}
\left[1-\frac{\sin\beta\phi_R-\sin\beta\phi_0}{\beta(\phi_R-\phi_0)}\right] + \cdots \;.
\end{equation}
Later we will comment on the meaning of this result in the UV
analysis of the ground state energy. On the other hand, comparing
the expansion for $k\to 1$ of ${\cal E}^{(3)}_{cl}(R)$ in the third regime with
\begin{equation}\label{mRIR}
mR\,=\,
-\log\left\{\frac{1-k^2}{16}\;\frac{1}{\tan\frac{\beta}{4}\phi_0\,\tan\frac{\beta}{4}\phi_R}\right\}+\cdots
\;,
\end{equation}
one obtains the large-$mR$ behaviour
\begin{equation}\label{eclIR}
{\cal E}^{(3)}_{cl}(R)\,=\, \frac{4m}{\beta^{2}}\,
\left(\,2-\cos\frac{\beta}{2}\phi_R-\cos\frac{\beta}{2}\phi_0\right)-
\frac{32m}{\beta^{2}}\,\tan\frac{\beta}{4}\phi_0\,\tan\frac{\beta}{4}\phi_R
\;e^{-mR}+\cdots \;.
\end{equation}
The first term of this expression is the classical limit of the boundary energy of the 
vacuum sector \cite{LMSS}, since it is the term that needs to be subtracted by choosing  
to normalise the energy to zero at $R\rightarrow \infty$. 

The classical description of the vacuum sector can be
completed by mentioning the existence of two particular cases in
which the three different regimes of $R$ are not needed. The first
is given by $\phi_{0} = \phi_{R}$, for which the whole range of $R$
is described by $\phi^{(3)}_{cl}(x)$ in (\ref{vacD}), since
$mR_{2} = 0$ in this situation.
The second case, defined by $\phi_{0}$ arbitrary and $\phi_{R} = 0$, can
be instead described by the antikink $\bar{\phi}^{(1)}_{cl}(x) =
\phi^{(1)}_{cl}(-x)$ alone, since $m R_{1} = \infty$ for these
values of the boundary parameters (note that $x_0$ and $R$ have
to be defined as opposite to the ones in (\ref{vacD})). As a consequence,
these two cases display a monotonic behaviour of the classical energy, 
whose UV and IR asymptotics, respectively, require a separate derivation,
which can be performed by simply adapting the above procedure.

Finally, it is also worth discussing an interesting feature which
emerges in the IR limit of the classical solution (\ref{vacD}).
As it can be seen from Fig.\,\ref{figvac}, by increasing $R$ the
static background is more and more localised closely to the
constant value $\phi(x)\equiv 0$ and this guarantees the
finiteness of the classical energy in the $R\to\infty$ limit,
given by the first term in (\ref{eclIR})\footnote{When
$|c_0|<|c_R|$, the same qualitative phenomenon occurs, but the
constant value is $\phi(x)\equiv \frac{2\pi}{\beta}$ in this
case.}. However, if the IR limit is performed directly on the
classical solution, we obtain one of the static
backgrounds\footnote{Obviously, the same function is obtained as
$\lim\limits_{R\to\infty}\bar{\phi}^{(1)}_{cl}(x)$, in the case
$\phi_R=0$ mentioned above.} studied in \cite{kormospalla}
$$
\phi^{(3)}_{cl}(x)\;{\mathrel{\mathop{\kern0pt\longrightarrow}
\limits_{R\to \infty}}}\;\frac{2}{\beta}\,
\arccos\left[\,\tanh\,m(x-x_{0}^{\infty})\right]\;,\qquad\text{with}\qquad
x_{0}^{\infty} \,=\, -\text{arctanh}(c_0)\;.
$$
The last expression tends to zero as $x\to\infty$ and consequently has classical energy
${\cal E}_{cl} = \frac{4m}{\beta^{2}}\,\left(\,1-\cos\frac{\beta}{2}\phi_0\right)$.
This phenomenon can be easily understood by noting that the minimum of
$\phi^{(3)}_{cl}(x)$ (which goes to zero in the IR limit), is placed
at $m\bar{x} = mx_{0} + \textbf{K}(k)$ (see Fig.\,\ref{figSGsol}) and this
point tends itself to infinity as $k\to 1$. Hence, the information about
the specific value of $\phi_R$ is lost when $R \rightarrow \infty$, i.e. only
the states with $\phi_R = 0$ survive in the IR limit.

\section{Semiclassical quantization on the strip}
\setcounter{equation}{0}

We will now perform the semiclassical quantization in the vacuum
sector, around the background (\ref{vacD}). Depending on the value
of $mR$, the stability equation (\ref{DHNstability}) takes the
form
\begin{equation}\label{stability1}
\left\{\frac{d^{2}}{d
\bar{x}^{2}}+k^{2}\left(\bar{\omega}^{2}+1\right)-2
k^{2}\,\textrm{sn}^{2}(\bar{x}-\bar{x}_{0},k)
\right\}\eta^{(1)}_{\bar{\omega}}(\bar{x}) \,=\, 0\;,\qquad
\text{with} \quad \bar{x}\,=\,\frac{m x}{k}\,,
\;\;\bar{\omega}\,=\,\frac{\omega}{m}\;,
\end{equation}
when $0<R<R_{1}$, and
\begin{equation}\label{stability2}
\left\{\frac{d^{2}}{d \bar{x}^{2}}+\bar{\omega}^{2}+1-2
k^{2}\,\textrm{sn}^{2}(\bar{x}-\bar{x}_{0},k)
\right\}\eta^{(2,3)}_{\bar{\omega}}(\bar{x}) \,=\, 0\;,\qquad
\text{with} \quad \bar{x}\,=\,m x\,,\;\;
\bar{\omega}\,=\,\frac{\omega}{m}\;,
\end{equation}
when $R_{1} < R < R_{2}$ and $R_{2} < R < \infty$.

%Equations (\ref{stability1}) and (\ref{stability2}) can be cast in
%the Lam\'e form with $N=1$, and for each $\bar{\omega}$ the
%solution is
%\begin{equation}
%\eta_{\bar{\omega}}(\bar{x})\,=\,A\,\eta_{a}(\bar{x})+B\,\eta_{-
%a}(\bar{x})\;,
%\end{equation}
%where
%\begin{equation}\label{lisol}
%\eta_{\pm a}(\bar{x}) \,=\,\frac{\sigma(\bar{x}-\bar{x}_{0}+i
%\textbf{K}' \pm a)}{\sigma(\bar{x}-\bar{x}_{0} + i
%\textbf{K}')}\;e^{\mp\,(\bar{x}-\bar{x}_{0}) \,\zeta(a)}\;,
%\end{equation}
%and
%\begin{equation}
%{\cal P}(a) \,= \,
%\begin{cases}
%\frac{2-k^{2}}{3}-k^{2}\bar{\omega}^{2} \qquad &\text{if}\quad 0<mR<mR_{1}\\
%\frac{2k^{2}-1}{3}-\bar{\omega}^{2} \qquad & \text{if}\quad
%mR_{1}<mR<\infty
%\end{cases}\;.
%\end{equation}

Equations (\ref{stability1}) and (\ref{stability2}) can be cast in
the Lam\'e form with $N=1$, which has been fully discussed in \cite{SGscaling}.
The only differences with the periodic case are the presence of a non-trivial
center of mass $x_0$ and the larger number of parameters entering the expression of 
the size $R$ of the system: 
these make more complicated the so--called \lq\lq quantization condition''
that determines the discrete eigenvalues, although they do not alter the general procedure
to derive it.

The boundary conditions (\ref{Dbc}), which translate in the requirement
\begin{equation} \eta_{\bar{\omega}}(0)
\,=\, \eta_{\bar{\omega}}(R) \,=\,0\,\;, \label{bceta}
\end{equation}
select in this case the following eigenvalues, all with multiplicity one,
\begin{equation}\label{omegaD}
\omega^{\text{vac}}_{n}(R)\,=\,
\begin{cases}
\omega^{(1)}_{n}(R)\qquad \text{for}\qquad 0<R<R_{1}\\
\omega^{(2)}_{n}(R)\qquad \text{for}\qquad R_{1}<R<R_{2}\\
\omega^{(3)}_{n}(R)\qquad \text{for}\qquad R_{2}<R<\infty
\end{cases}\;,
\end{equation}
where
\begin{eqnarray}\label{omegaDexplicit}
&\omega^{(1)}_{n}(R)&=\;\frac{m}{k}\sqrt{\frac{2-k^{2}}{3}-{\cal P}(i y_{n})} \;\;,\nonumber\\
&\omega^{(2,3)}_{n}(R)&=\;m\sqrt{\frac{2k^{2}-1}{3}-{\cal P}(i
y_{n})}  \;\;,\nonumber
\end{eqnarray}
and the $y_{n}$'s are defined through the \lq\lq quantization condition''
\begin{equation}\label{quantcond}
2\bar{R}\,i\,\zeta(i y_{n})+i\,\log\left[\frac{\sigma(-\bar{x}_{0}
+ i \textbf{K}'+i y_{n})\,\sigma(\bar{R}-\bar{x}_{0} + i
\textbf{K}'-i y_{n})}{\sigma(-\bar{x}_{0} + i \textbf{K}'-i
y_{n})\,\sigma(\bar{R}-\bar{x}_{0} + i \textbf{K}'+i
y_{n})}\right]\,=\,\,2 n \pi\;,\qquad n=1,2,...
\end{equation}
This equation comes from the consistency condition associated to
the boundary values
$$
\begin{cases}
D_+ \,\eta_{a}(0)+D_-\,\eta_{-a}(0) = 0 \,\,\,\,\;,\\
D_+ \,\eta_{a}(R)+D_-\,\eta_{-a}(R) = 0 \,\,\,,
\end{cases}
$$
where $\eta_{\pm a}$ are the two linearly independent solutions of
the Lam\'e equation which are used to build the general solution
$\eta(x)=D_+\,\eta_{a}(x)+D_-\,\eta_{- a}(x)$ (see
\cite{SGscaling} and \cite{whit} for details).

\begin{figure}[h]
\footnotesize \psfrag{omega1}{$\omega_{i}^{\text{vac}}/m$}
\psfrag{ell}{$mR$}
\begin{center}
\psfig{figure=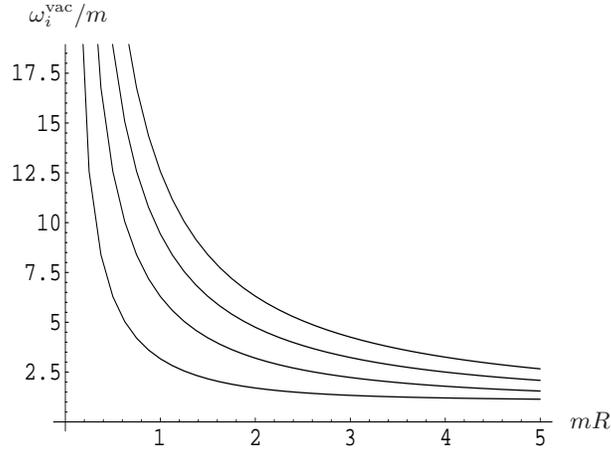,height=6cm,width=8cm}
\end{center}
\caption{The first few energy levels (\ref{omegaD}) for
$\beta\phi_{0}=1$ and $\beta\phi_{R}=2$.} \label{figomegai}
\end{figure}

As it can be seen directly from (\ref{DHNlevels}), the frequencies (\ref{omegaD})
are nothing else but the energies of the excited states with respect
to the ground state $E^{\text{vac}}_0(R)$. They can be easily determined from the above
equations and their behaviour, as functions of $R$, is shown in Fig.\,\ref{figomegai}.

As in the periodic case \cite{SGscaling}, a more explicit expression
for the energy levels  (\ref{omegaD}) can be obtained by expanding them
for small or large values of $mR$. The UV expansion, for instance,
can be performed extracting from (\ref{quantcond}) a small-$k$ expansion for $y_{n}$,
inserting it in
(\ref{omegaD}), and finally comparing the result order by order with (\ref{mRUV}).
Exploiting the several properties
of Weierstrass functions which follow from their relation with $\theta$--functions
(see for instance \cite{whit}), one gets
$$
y_n\,=\,\text{arctanh}\,\frac{f}{2n\pi}
+\frac{k^2}{4}\left\{\text{arctanh}\,\frac{f}{2n\pi}+
\,s\;\,\frac{2n\pi(4n^2\pi^2-3f^2)}{(4n^2\pi^2-f^2)^2} \right\} + \cdots \;,
$$
and
$$
\omega_n^{(1)}\,=\,\frac{m}{k}\,\frac{2n\pi}{f}\left\{1-\frac{k^2}{4}
\left[1+\frac{s}{f}-\frac{2 f s}{4n^2\pi^2-f^2}\right] + \cdots
\right\}\;,
$$
where we have introduced the compact notation $f\equiv\beta(\phi_R-\phi_0)$,
$s\equiv(\sin\beta\phi_R-\sin\beta\phi_0)$. This leads to the UV expansion
\begin{equation}\label{omegaDUV}
\omega_n^{(1)}(R)\,=\,\frac{n\pi}{R}+m^2 R \,\;\frac{s}{f}\;
\frac{2n\pi}{4n^2\pi^2-f^2} + \cdots
\end{equation}

In order to complete the above analysis and obtain the reference value of the energy levels,
i.e. the ground state energy $E^{\text{vac}}_0(R)$ of the vacuum sector, we need the classical
energy (\ref{eclD}) and the sum on the stability frequencies given in (\ref{omegaD}), i.e.
\begin{equation}\label{grstatevac}
 E^{\text{vac}}_0(R) \,=\, {\cal E}^{\text{vac}}_{cl}(R) +
\frac{1}{2}\,\sum_{n=1}^{\infty} \omega^{\text{vac}}_n(R)\;.
\end{equation}
The above series is divergent and its regularization has to be
performed by subtracting to it a mass counterterm and the
divergent term coming from the infinite volume limit -- a
procedure that is conceptually analogous to the one discussed in
\cite{SGscaling} for the periodic case and therefore it is not
repeated here. Furthermore, as already mentioned, equation
(\ref{grstatevac}) can be made more explicit by expanding it for
small or large values of $mR$. Here, for simplicity, we limit 
ourselves to the discussion of the leading $1/R$ term in the UV 
expansion since it does 
not receive contributions from the counterterm and therefore it 
can be simply regularised by using the Riemann $\zeta$--function 
prescription (see \cite{SGscaling} for a detailed discussion). The 
higher terms, instead, require a technically more complicated
regularization, although equivalent to the one presented in
\cite{SGscaling}.

The UV behaviour of the ground state energy is dominated by
\begin{equation}\label{e0UV}
E^{\text{vac}}_0(R) \,=\,
\frac{\pi}{R}\left[\frac{1}{2\pi}\left(\phi_R-\phi_0\right)^2
- \frac{1}{24}\right] + \cdots
\end{equation}
where the coefficient $-1/24$ comes from the regularization of the leading term
in the series of frequencies (\ref{omegaDUV}), while the first term simply comes
from the expansion of the classical energy (\ref{eclUV}). It is easy to see that
the above expression correctly reproduces the expected ground state energy for
the gaussian Conformal Field Theory (CFT) on a strip of width $R$ with Dirichlet
boundary conditions \cite{cardy,saleurlect}.

Finally, it is simple to check that also the excited energy levels display
the correct UV behaviour, being expressed as
\begin{equation}\label{eiUV}
E^{\text{vac}}_{\{k_{n}\}}(R)\, = \,
\frac{\pi}{R}\,\left[\frac{1}{2\pi}\left(\phi_R-\phi_0\right)^2 +
\sum\limits_{n}k_{n}\,n-\frac{1}{24}\right] + \cdots
\end{equation}

\section{The kink sector: $Q=1$}
\setcounter{equation}{0}

In discussing the kink sector we can restrict to
$n_0=0\,,\;n_R=1$, since all other cases, as well as the antikink
sector with $Q=-1$, are described by straightforward
generalizations of the following formulas.

\begin{figure}[h]
\begin{tabular}{p{8cm}p{8cm}}

\footnotesize

\psfrag{phi}{$\beta\phi_{cl}^{\text{kink}}$} \psfrag{ell}{$x$}

\psfig{figure=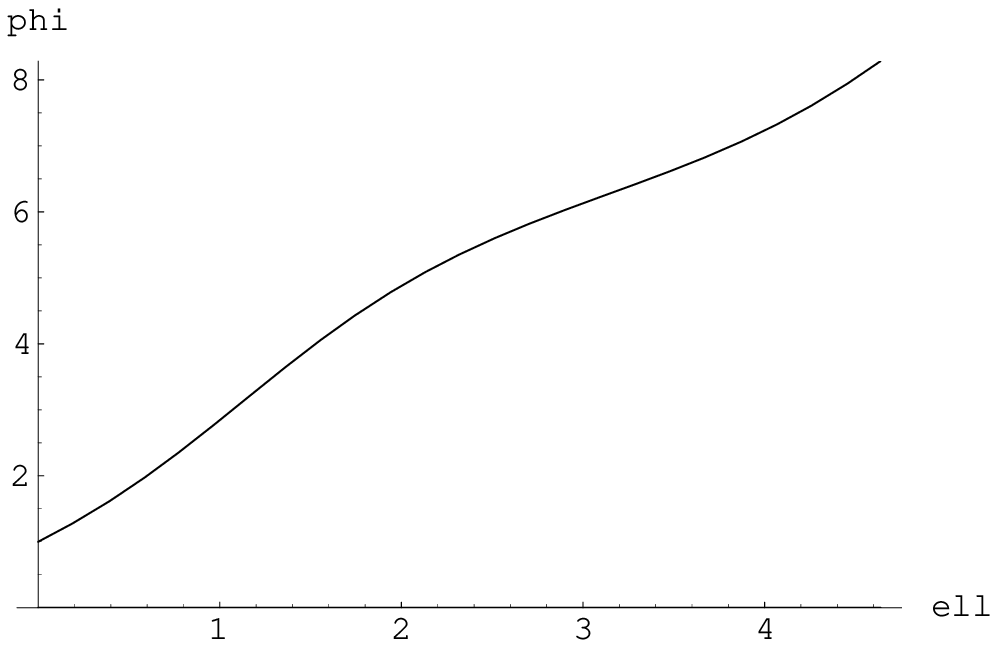,height=5cm,width=7cm}&

\footnotesize

\psfrag{phi}{$\beta\phi_{cl}^{\text{kink}}$} \psfrag{ell}{$x$}

\psfig{figure=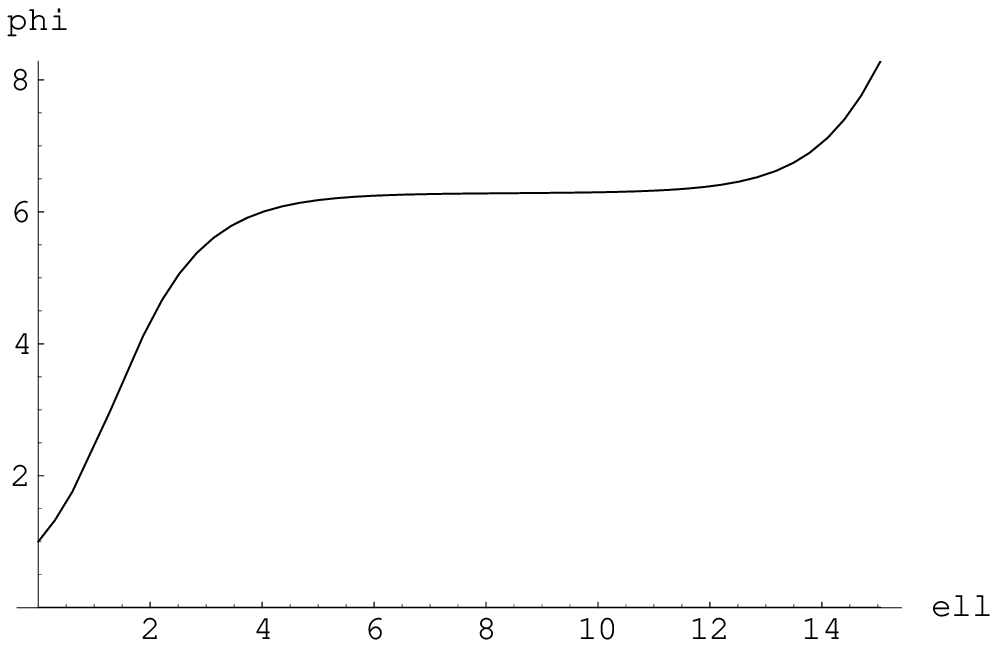,height=5cm,width=7cm}
\end{tabular}
\caption{Classical solution (\ref{kinkD}) at some values of $R$,
in the case $\beta\phi_{0}=1$ and $\beta\phi_{R}=2$.}
\label{figkink}
\end{figure}

The classical solution can be now expressed only in terms of the
function $\phi^{(+)}_{cl}(x)$ as
\begin{equation}\label{kinkD}
\phi_{cl}^{\text{kink}}(x)\,=\,\phi^{(+)}_{cl}(x)\qquad
\text{with}\qquad
\begin{cases}
m x_{0} = -k\,F\left(\frac{\beta}{2}\phi_{0}-\frac{\pi}{2},k\right)\\
m R = k\left[2
\textbf{K}(k)+F\left(\frac{\beta}{2}\phi_{R}-\frac{\pi}{2},k\right)-
F\left(\frac{\beta}{2}\phi_{0}-\frac{\pi}{2},k\right)\right]\\
0 < k < 1
\end{cases}\;,
\end{equation}
since in this case the whole range $0 < m R < \infty$ is spanned
by varying $k$ in $[0,1]$. This can be intuitively understood by
looking at the behaviour of (\ref{kinkD}) in Fig.\,\ref{figkink}.

As a consequence, the classical energy and the stability
frequencies of this sector can be obtained from ${\cal
E}_{cl}^{(1)}$ and $\omega_{n}^{(1)}$ of the vacuum (given
respectively in eq.\,(\ref{eclD}) and (\ref{omegaD})), by simply
replacing $\phi_{R}\to \phi_{R}+\frac{2\pi}{\beta}$. The leading
UV behaviour of the energy levels in this sector, given by
\begin{equation}\label{kinkeiUV}
E^{\text{kink}}_{\{k_{n}\}}(R)\, = \,
\frac{\pi}{R}\,\left[\frac{1}{2\pi}
\left(\left(\phi_R-\phi_0\right)+\frac{2\pi}{\beta}\,Q\right)^2 +
\sum\limits_{n}k_{n}\,n-\frac{1}{24}\right] + \cdots
\end{equation}
with $Q=1$, correctly matches the CFT prediction.

\begin{figure}[h]
\footnotesize \psfrag{ec}{$\beta^{2}{\cal
E}^{\text{kink}}_{cl}/m$} \psfrag{ell}{$mR$}
\begin{center}
\psfig{figure=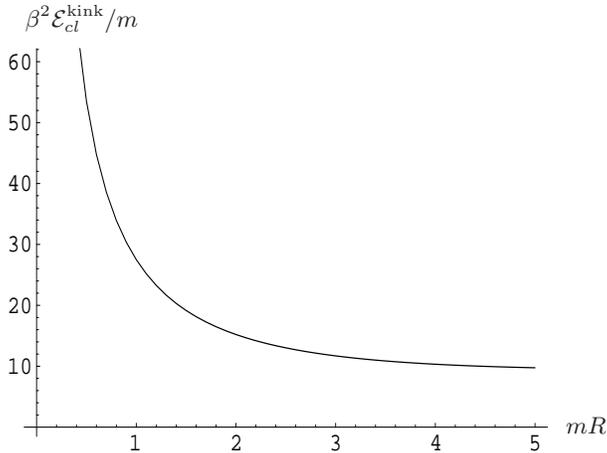,height=6cm,width=8cm}
\end{center}
\caption{Classical energy in the $Q=1$ kink sector for $\beta\phi_{0}=1$
and $\beta\phi_{R}=2$.} \label{figeclkink}
\end{figure}

The only result which cannot be directly extracted from the vacuum
sector analysis is the IR asymptotic behaviour of the classical
energy, since now the $k\to 1$ limit has to be performed on ${\cal
E}^{(1)}_{cl}$. We have in this case
\begin{equation}\label{mRIRkink}
mR\,=\,
-\log\left\{\frac{1-k^2}{16}\;\frac{\tan\frac{\beta}{4}\phi_0}
{\tan\frac{\beta}{4}\phi_R}\right\}+\cdots
\;,
\end{equation}
which leads to
\begin{equation}\label{eclIRkink}
{\cal E}^{(1)}_{cl}(R)\,=\, \frac{4m}{\beta^{2}}\,
\left(\,2-\cos\frac{\beta}{2}\phi_R+\cos\frac{\beta}{2}\phi_0\right)+
\frac{32m}{\beta^{2}}\,\frac{\tan\frac{\beta}{4}\phi_R}
{\tan\frac{\beta}{4}\phi_0}\;e^{-mR}+\cdots
\;.
\end{equation}
Analogously to the vacuum sector, the first term of this expression 
is related to the classical limit of the boundary energy in the one--kink 
sector. Notice that, differently from the vacuum case, where the asymptotic 
IR value of the classical energy was approached from below (see
(\ref{eclIR})), the coefficient of the exponential correction has
now positive sign, in agreement with the monotonic behaviour of
the classical energy shown in Fig.\,\ref{figeclkink}.

When $R\to\infty$, a mechanism analogous to the one discussed for
the vacuum also takes place here: the classical energy is finite
for any value of $\phi_R$, but since $\phi_{cl}^{\text{kink}}(x)$
assumes the value $\frac{2\pi}{\beta}$ at $m\bar{x} = mx_{0} +
k\textbf{K}(k)$ (see Fig.\,\ref{figSGsol}), a point which tends to
infinity as $k\to 1$, only the states with $\phi_R = 0$ survive in
this limit. 

\begin{figure}[h]
\begin{tabular}{p{8cm}p{8cm}}

\footnotesize

\psfrag{phi}{$\beta\phi_{cl}$} \psfrag{ell}{$x$}

\psfig{figure=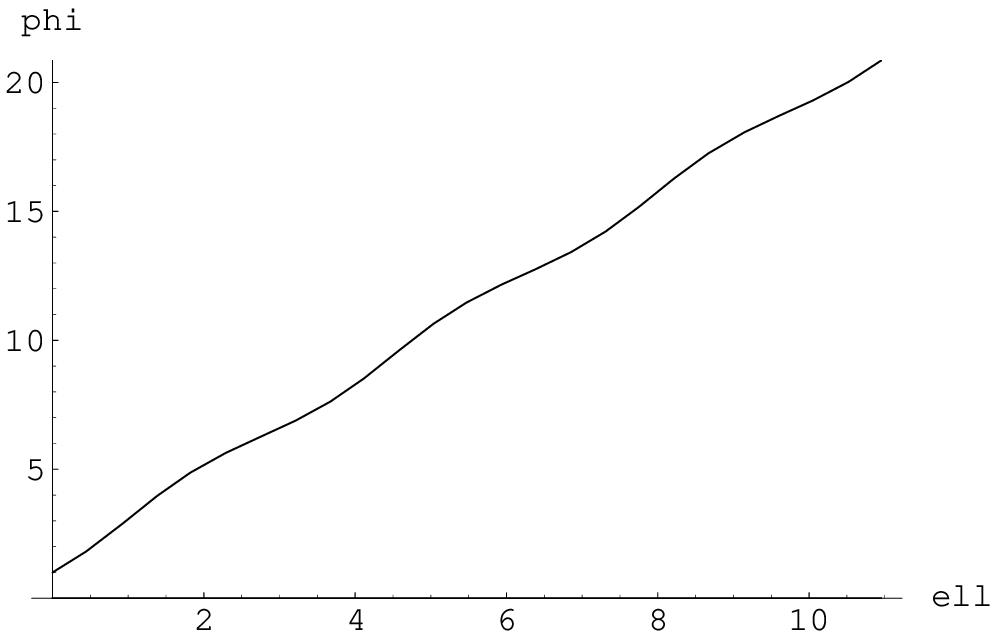,height=5cm,width=7cm}&

\footnotesize

\psfrag{phi}{$\beta\phi_{cl}$} \psfrag{ell}{$x$}

\psfig{figure=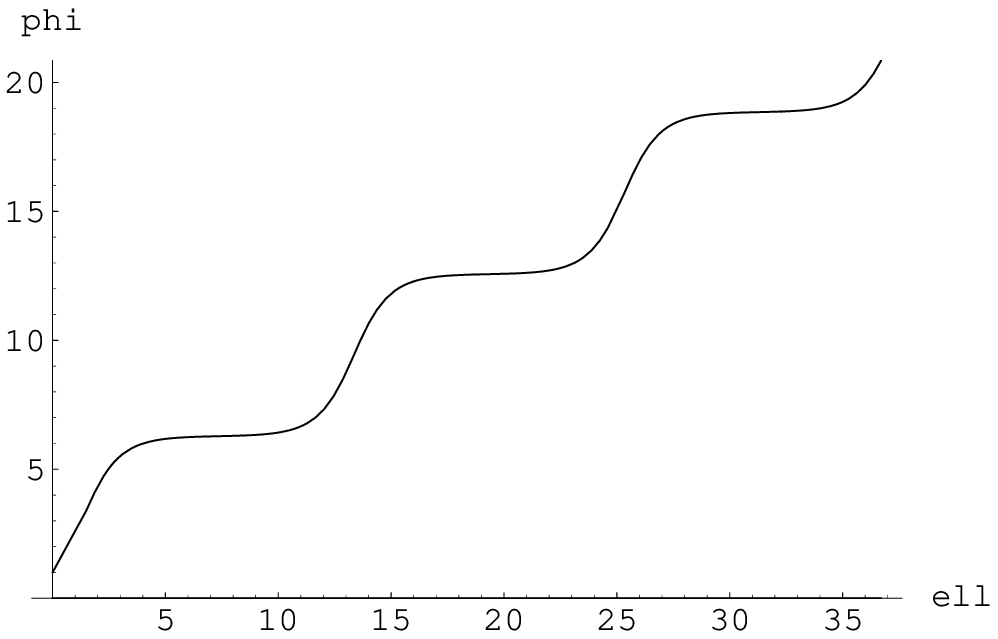,height=5cm,width=7cm}
\end{tabular}
\caption{Classical solution in the $Q=3$ sector ($n_0=0$, $n_R=3$) at some values of $R$,
in the case $\beta\phi_{0}=1$ and $\beta\phi_{R}=2$.}
\label{fig3kink}
\end{figure}

It is worth noticing that $\phi^{(+)}_{cl}(x)$ can be also 
used to satisfy, at finite values of $R$, Dirichlet b.c. in sectors with 
arbitrary topological charge (see Fig.\,\ref{fig3kink}), giving rise to the correct UV 
behaviour (\ref{kinkeiUV}) with $Q = n_R - n_0$. However, since $\phi^{(+)}_{cl}(x)$ 
always assumes the value $\frac{2\pi}{\beta}(n_0+1)$ at $m\bar{x} = mx_{0} +
k\textbf{K}(k)$, which is once again the point going to infinity when $k\to 1$, 
in the IR limit it can only correspond to $Q = 1$. This result seems 
natural though, since in infinite volume, static classical solutions 
can only describe those sectors of the theory with $Q=0,\pm 1$, while time--dependent 
ones are needed for higher values of $Q$. 
Hence, in the topological sectors 
with $|Q| > 1$ the space of states will contain, at classical level, the
time--dependent backgrounds, defined for any value of $R$ (which
are not discussed here), plus the static ones of the form
$\phi^{(+)}_{cl}(x)$, which however disappear from the spectrum as
$R\to\infty$.

\section{Conclusions}

In this paper we have presented the semiclassical energy levels of
a quantum field theory on a strip geometry. Our analysis builds on
and extends the semiclassical quantization of a field theory on a
finite geometry introduced in \cite{SGscaling}. The example
discussed here is the Sine--Gordon model with Dirichlet boundary
conditions at both edges of the strip. The semiclassical approach
provides analytic and non--perturbative expressions for the energy
levels, valid for arbitrary values of the size $R$ of the system,
which permit to link the IR data on the half-line with
the UV conformal data of boundary CFT at $c=1$.

In comparison with a cylinder geometry, an interesting new feature
of the quantum field theory defined on a strip consists in a
non--trivial (and non--perturbative) semiclassical description of
its vacuum sector. Therefore, we have discussed in detail the
classical solutions and energy levels in the $Q=0$ case, together
with the $Q=1$ that can also be described by static backgrounds.
It should be mentioned, however, that the semiclassical methods
\cite{DHN} are not restricted to static backgrounds only. As in infinite
volume, a complete description of the theory in all sectors
requires also the study of non--perturbative time--dependent solutions.

Finally, it is worth noticing that the method used here has
natural and direct extension to other quantum field theories with
various kinds of boundary conditions.

\vspace{0.5cm}

\begin{flushleft}\large
\textbf{Acknowledgments}
\end{flushleft}
This work was partially supported by the Italian COFIN contract
\lq\lq Teoria dei Campi, Meccanica Statistica e Sistemi
Elettronici'' and by the European Commission TMR programme
HPRN-CT-2002-00325 (EUCLID). One of us (GM) would like to thank
the Laboratoire de Physique Theorique in Jussieu (Paris) for the
kind hospitality during the last stage of this work.

\vspace{5mm}

\noindent {\large {\em Note added.}} In completing this
work, it has appeared the paper \cite{hungstrip} which 
partially overlaps with ours.


\vspace{5mm}

\begin{thebibliography}{1}
\bibitem{DHN} R.F. Dashen, B. Hasslacher and A. Neveu, Phys. Rev. D 10
(1974) 4130; Phys. Rev. D 11 (1975) 3424.

\bibitem{goldstone} J. Goldstone and R. Jackiw, Phys. Rev. D 11 (1975)
1486.

\bibitem{finvolff} G. Mussardo, V. Riva and G. Sotkov,
Nucl. Phys. B 670 (2003) 464.

\bibitem{dsgmrs} G. Mussardo, V. Riva and G. Sotkov,
Nucl. Phys. B 687 (2004) 189.

\bibitem{SGscaling} G. Mussardo, V. Riva and G. Sotkov,
\textit{Semiclassical Scaling Functions of Sine--Gordon Model},
hep-th/0405139.

\bibitem{ghoshzam} S. Ghoshal and A.B. Zamolodchikov, Int. J. Mod. Phys. A9 (1994) 3841,
Erratum-ibid. A9 (1994) 4353.

\bibitem{mattdorey} P. Mattsson and P. Dorey, J. Phys. A33 (2000) 9065.

\bibitem{BPTT} Z. Bajnok, L. Palla, G. Takacs and  G.Z. Toth,
Nucl. Phys. B 622 (2002) 548.

\bibitem{SSW}  H. Saleur, S. Skorik and N.P. Warner, Nucl. Phys. B 441 (1995) 421.

\bibitem{corr} E. Corrigan and G.W. Delius, J. Phys. A 32 (1999) 8601;

E. Corrigan and A. Taormina, J. Phys. A 33 (2000) 8739.

\bibitem{kormospalla} M. Kormos and L. Palla, J. Phys. A35 (2002) 5471.

\bibitem{LMSS} A. LeClair, G. Mussardo, H. Saleur and S. Skorik,
Nucl. Phys. B 453 (1995) 581.

\bibitem{SS} S. Skorik and H. Saleur, J. Phys. A 28 (1995) 6605.

\bibitem{cauxsal} J.S. Caux, H. Saleur and F. Siano, Phys. Rev. Lett.88 (2002) 106402;
Nucl. Phys. B 672 (2003) 411.

\bibitem{leerim} T. Lee and C. Rim, Nucl. Phys. B 672 (2003) 487.

\bibitem{rim} C. Rim, \textit{Boundary massive Sine-Gordon model at the free Fermi
limit and RG flow of Casimir energy}, hep-th/0405162.

\bibitem{ahnnep} C. Ahn and R.I. Nepomechie,  Nucl. Phys. B 676 (2004) 637.

\bibitem{rav} C. Ahn, M. Bellacosa and F. Ravanini, \textit{Excited states NLIE
for Sine-Gordon model in a strip with Dirichlet boundary conditions}, hep-th/0312176.


\bibitem{takoka} K. Takayama and M. Oka, Nucl. Phys. A 551 (1993) 637.

\bibitem{GRA} I.S. Gradshteyn and I.M. Ryzhik,
\textit{Table of integrals, series, and products}, Academic Press,
New York (1980).

\bibitem{whit} E.T.Whittaker and G.N.Watson, \textit{A course of modern analysis},
Cambridge, Cambridge University Press, 1927.



\bibitem{cardy} H.W.J. Blote, J.L. Cardy and M.P. Nightingale,
Phys. Rev. Lett. 56 (1986) 742;

J.L. Cardy, Nucl. Phys. B 270 (1986) 186; Nucl. Phys. B 275 (1986) 200;
\textit{Conformal invariance and statistical mechanics}, Les Houches 1988,
North Holland, Amsterdam.

\bibitem{saleurlect} H. Saleur, \textit{Lectures on nonperturbative field theory
and quantum impurity problems}, Les Houches 1998,
North Holland, Amsterdam, cond-mat/9812110.

\bibitem{hungstrip} Z. Bajnok, L. Palla and G. Takacs,
\textit{(Semi)classical analysis of sine-Gordon theory
on a strip}, hep-th/0406149.

\end{thebibliography}
\end{document}